\documentclass[prl,amsmath,twocolumn,amssymb,superscriptaddress,floatfix]{revtex4-1} 
\usepackage[colorlinks]{hyperref}
\usepackage{graphicx}
\usepackage{amsmath, braket}
\usepackage{amssymb}
\usepackage{natbib}
\usepackage[normalem]{ulem} 
\usepackage[dvipsnames]{xcolor}
\usepackage[version=4]{mhchem}
\usepackage[english]{babel}
\usepackage[parse-numbers=false]{siunitx} 
\usepackage{xspace} 

\newcommand{\tas}{\ce{TaS2}\xspace}
\newcommand{\nbse}{\ce{NbSe2}\xspace}

\newcommand{\musq}{$\SI{}{\micro m^2\,}$}

\begin{document}

\title{Superconducting Contact and Quantum Interference \\ Between Two-Dimensional van der Waals and Three-Dimensional Conventional Superconductors}

\author{Michael~R.~Sinko}
\author{Sergio~C.~de~la~Barrera} 
\affiliation{Department of Physics, Carnegie Mellon University, Pittsburgh, PA 15213}
\author{Olivia~Lanes}
\affiliation{Department of Physics and Astronomy, University of Pittsburgh, Pittsburgh, PA 15260}
\author{Kenji~Watanabe}
\author{Takashi~Taniguchi}
\affiliation{Advanced Materials Laboratory, National Institute for Materials Science, Tsukuba, Ibaraki 305-0044, Japan}
\author{Susheng~Tan}
\affiliation{Department of Electrical and Computer Engineering, University of Pittsburgh, Pittsburgh, PA 15260}
\affiliation{Petersen Institute of Nanoscience and Engineering, University of Pittsburgh, Pittsburgh, PA 15260}
\author{David~Pekker}
\affiliation{Department of Physics and Astronomy, University of Pittsburgh, Pittsburgh, PA 15260}
\author{Michael~Hatridge} \email{hatridge@pitt.edu}
\affiliation{Department of Physics and Astronomy, University of Pittsburgh, Pittsburgh, PA 15260}
\author{Benjamin~M.~Hunt} \email{bmhunt@andrew.cmu.edu}
\affiliation{Department of Physics, Carnegie Mellon University, Pittsburgh, PA 15213}

\begin{abstract}
Two-dimensional (2D) transition-metal dichalcogenide superconductors have unique and desirable properties for integration with conventional superconducting circuits. 
However, fully superconducting contact must be made between the 2D material and three-dimensional (3D) superconductors in order to employ the standard microwave drive and readout of qubits in such circuits. Here, we present a method for creating zero-resistance contacts between 2D \nbse and 3D aluminum that behave as Josephson junctions (JJs) with large effective areas compared to 3D-3D JJs. We present a model for the supercurrent flow in a 2D-3D superconducting structure by numerical solution of the Ginzburg-Landau equations and find good agreement with experiment. These results demonstrate 
a crucial step towards a new generation of hybrid superconducting quantum circuits.
\end{abstract}

\texorpdfstring{\maketitle}{\maketitle}

Superconducting circuits are ubiquitous in the fields of quantum information and quantum sensing.
Typically fabricated by deposition of metallic superconductors such as Al, Nb, or NbTiN, most state-of-the-art superconducting qubits use Josephson junctions whose tunnel barriers are created by in-situ oxidation of aluminum electrodes \cite{devoret_superconducting_2013,devoret_implementing_2004}. Despite well-controlled processes for creating Josephson junctions, the oxide barrier thickness varies on the atomic scale, leading to highly non-uniform supercurrent distributions \cite{zeng_direct_2015}. Oxide barriers also age as oxygen atoms diffuse out from the oxide, changing the normal state resistance of the junction and hence its critical current \cite{ambegaokar_tunneling_1963}; similar effects result from the absorption or desorption of other atoms or molecules \cite{zeng_direct_2015,pop_fabrication_2012,scherer_effect_2001}. 
Despite these limitations, Josephson junction-based superconducting circuits are at the vanguard of quantum computing, used as state-of-the-art qubits but also employed in readout cavities, filters, amplifiers, and circulators. This is enabled by varying their size by several orders of magnitude to engineer desired nonlinear Hamiltonians, and by embedding junctions in a superconducting loop to form Superconducting QUantum Interference Devices (SQUIDs), as well as related variants such as Superconducting Low-inductance Undulatory Galvanometers (SLUGs), Superconducting Nonlinear Asymmetric Inductive eLements (SNAILs), and Josephson Ring Modulators (JRMs) \cite{bergeal_analog_2010,frattini_optimizing_2018,clarke_squid_2004,devoret_superconducting_2013,devoret_implementing_2004}.

\begin{figure}[hb!]
  \begin{center}
\includegraphics[width=86mm]{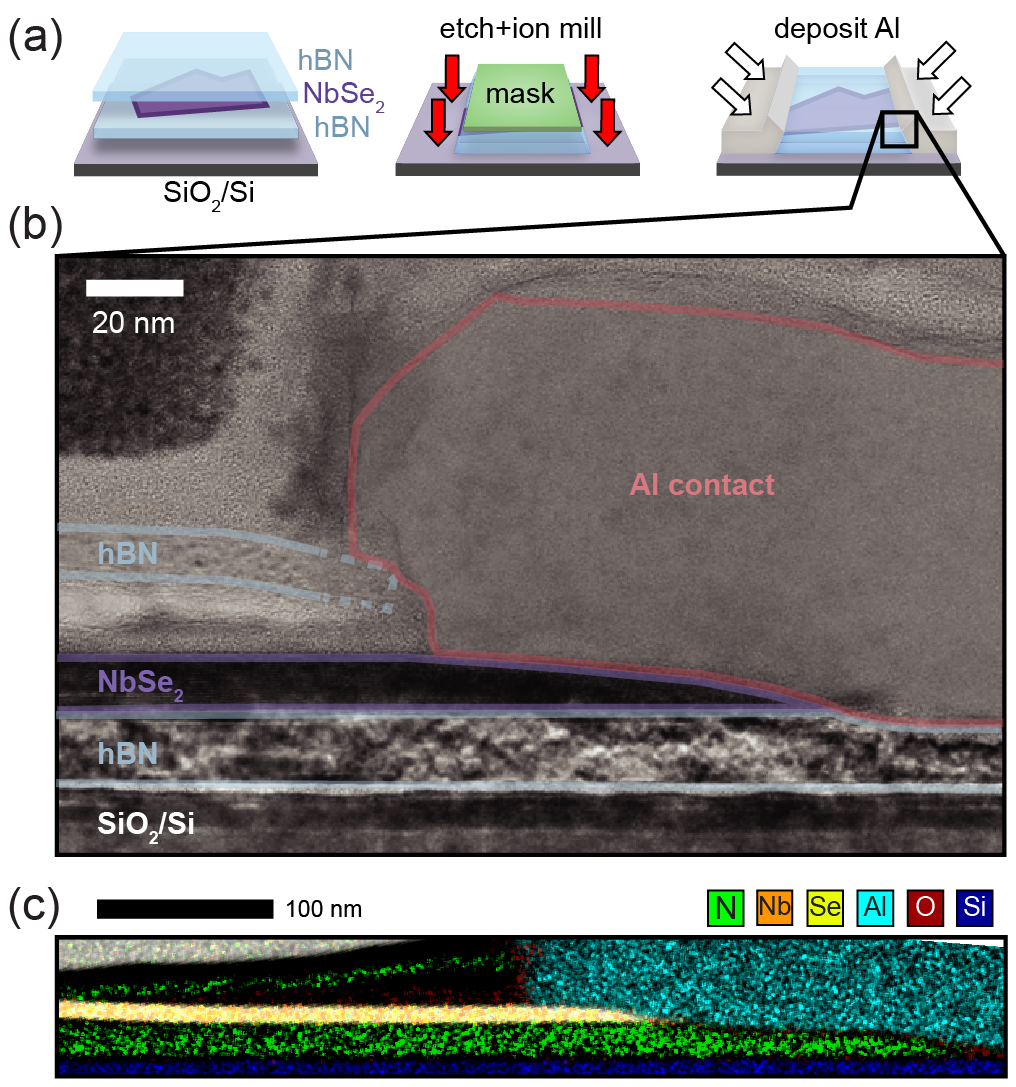} 
\caption{ \textbf{Contact method and cross-sectional transmission electron microscopy (TEM).}
\textbf{(a)} Thin NbSe$_2$ is encapsulated by hBN.  A reactive-ion etch exposes the NbSe$_2$ edge. Following an in-situ argon ion mill, contact is made using angled evaporation of aluminum with no adhesion layer.  \textbf{(b)} Cross-sectional TEM of NbSe$_2$-Al interface in device F.   \textbf{(c)} False-color EDS map of the \nbse-Al contact in device G.  A slight delamination of the top hBN near the contact is visible, as in (b). Minimal oxygen is observed at the \nbse-Al contact.}
\label{fig-TEM}
  \end{center}
\end{figure}

\begin{figure}[ht!]
  \begin{center}
\includegraphics[width=85mm]{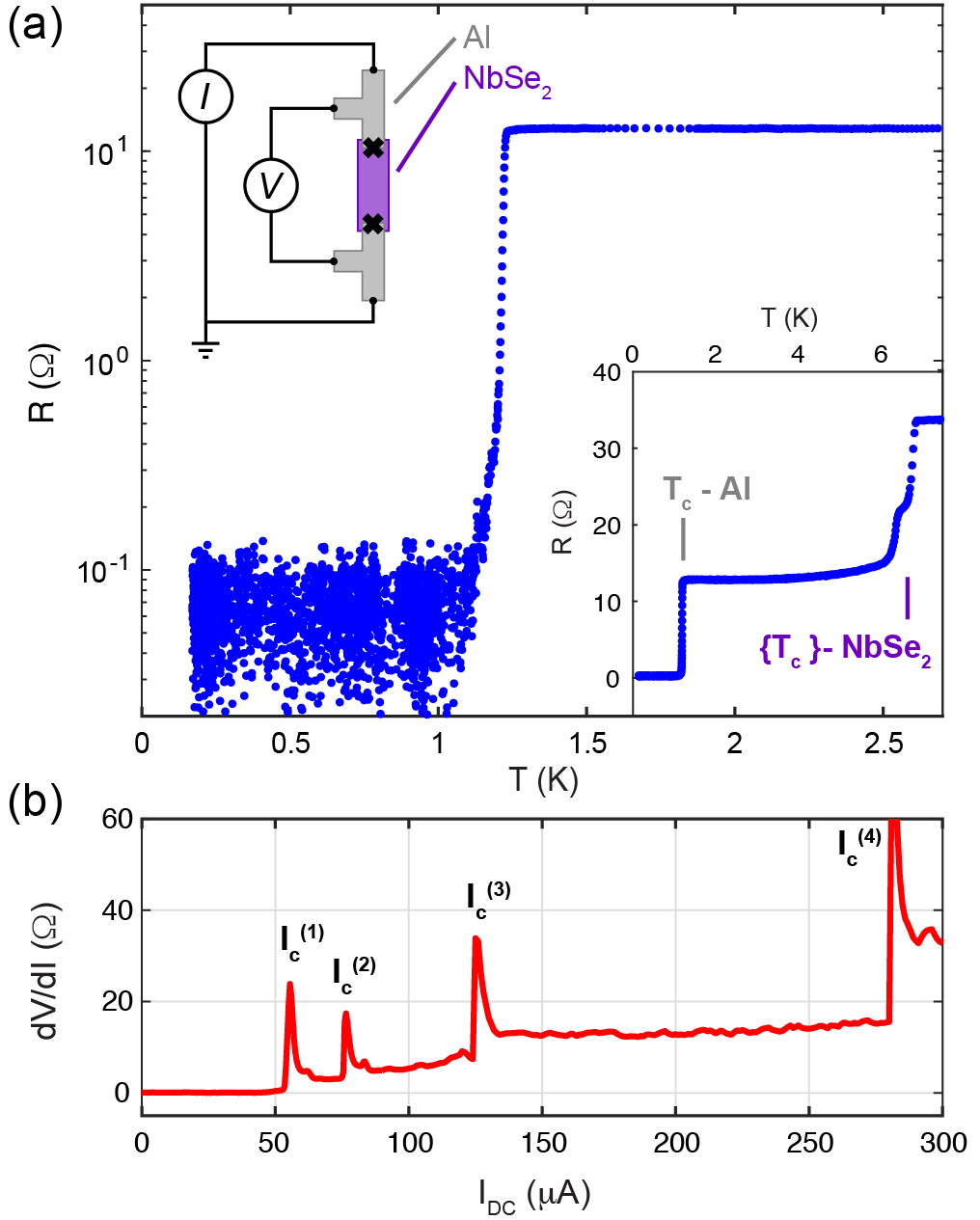} 
\caption{ \textbf{Zero-resistance contact between few-layer NbSe$_2$ and aluminum.}
\textbf{(a)} $R(T)$ for sample F, showing a zero-resistance state. Inset (bottom-right): linear plot of $R(T)$ showing the Al and two \nbse transitions. Inset (top-left): 4-pt measurement setup for measuring the contact resistances of the Al/\nbse contacts. 
\textbf{(b)} Differential resistance ($dV/dI$) vs.~DC current $I_{\rm DC}$ of same device, showing four DC critical currents.}
\label{fig-20}
  \end{center}
\end{figure}

Two-dimensional van der Waals (vdW) superconductors, such as 2H$_a$-\nbse and 2H$_a$-\tas \cite{barrera_tuning_2018, navarro-moratalla_enhanced_2016} (hereafter \nbse and \tas), offer complementary properties that leverage their atomic thinness and unique magnetic properties. For example, kinetic inductance $L_{\rm{K}}$, which results from the inertia of the charge carriers and is inversely proportional to cross-sectional area, can be quite large in a vdW superconductor because the atomic thinness minimizes the cross-sectional area for a given channel width. A large kinetic inductance is desirable when fabricating superinductors for superconducting circuits like fluxonium qubits, or inductive shunts \cite{vissers_frequency-tunable_2015,annunziata_tunable_2010}. Additionally, \nbse and \tas can be readily incorporated in heterostructures in which hexagonal boron nitride (hBN) or the semiconducting transition-metal dichalcogenides (TMDs) could replace oxides as stable and atomically-flat tunnel barriers\cite{wang_coherent_2018,lee_two-dimensional_2019,kim_strong_2017,kim_van_2016}. They also possess strong spin-orbit coupling, which leads to exceptionally high in-plane upper critical fields in few-layer devices \cite{xi_evidence_2016,barrera_tuning_2018}.

In order to take advantage of the extraordinary properties of 2D vdW superconductors, the most promising pathway is to integrate components built from 2D superconductors into conventional 3D superconducting circuits. The \textit{sine qua non} of creating such a hybrid circuit is to create reliable, transparent and robust superconducting contact between the 2D vdW and 3D conventional materials, minimizing dissipation and allowing the use of standard microwave drive and readout techniques. In this work, we present such a method, demonstrating zero-resistance contacts between 2D \nbse and 3D aluminum. In addition, we study the magnetic-flux response of devices having both two-terminal (3D-2D-3D) and SQUID geometries. With the aid of numerical solutions of the Ginzburg-Landau equations, we elucidate how the response of 2D-3D superconducting devices to magnetic field depends strongly on the geometry of the 2D flake due to the gradual variation of the screening currents in two dimensions. 

We begin by describing our fabrication methods, shown in Fig.~\ref{fig-TEM}(a).  Using exfoliation and standard dry-transfer techniques for stacking van der Waals materials inside a nitrogen-filled glove box \cite{tsen_nature_2016}, we encapsulate few-layer \nbse above and below with hexagonal boron nitride (hBN). We make contact to the \nbse using evaporated aluminum with no intermediate adhesion layer, using a variation of the ``edge contact'' method \cite{wang_one-dimensional_2013}. Briefly, we expose the edge of the \nbse by reactive-ion etching through all three layers of the hBN-\nbse-hBN stack. We then transfer the stack immediately into an evaporation chamber for an argon ion mill process, to clean the exposed cross section of the stack, after which we tilt the sample \textit{in situ} in the appropriate orientation to evaporate Al onto the exposed \nbse edge and surface (Fig.~\ref{fig-TEM}(b) shows a cross-sectional TEM image of the result in one of our devices). We have successfully made devices using both single- and double-angle evaporation.


\begin{figure*}[ht!]
  \begin{center}
\includegraphics[width=170mm]{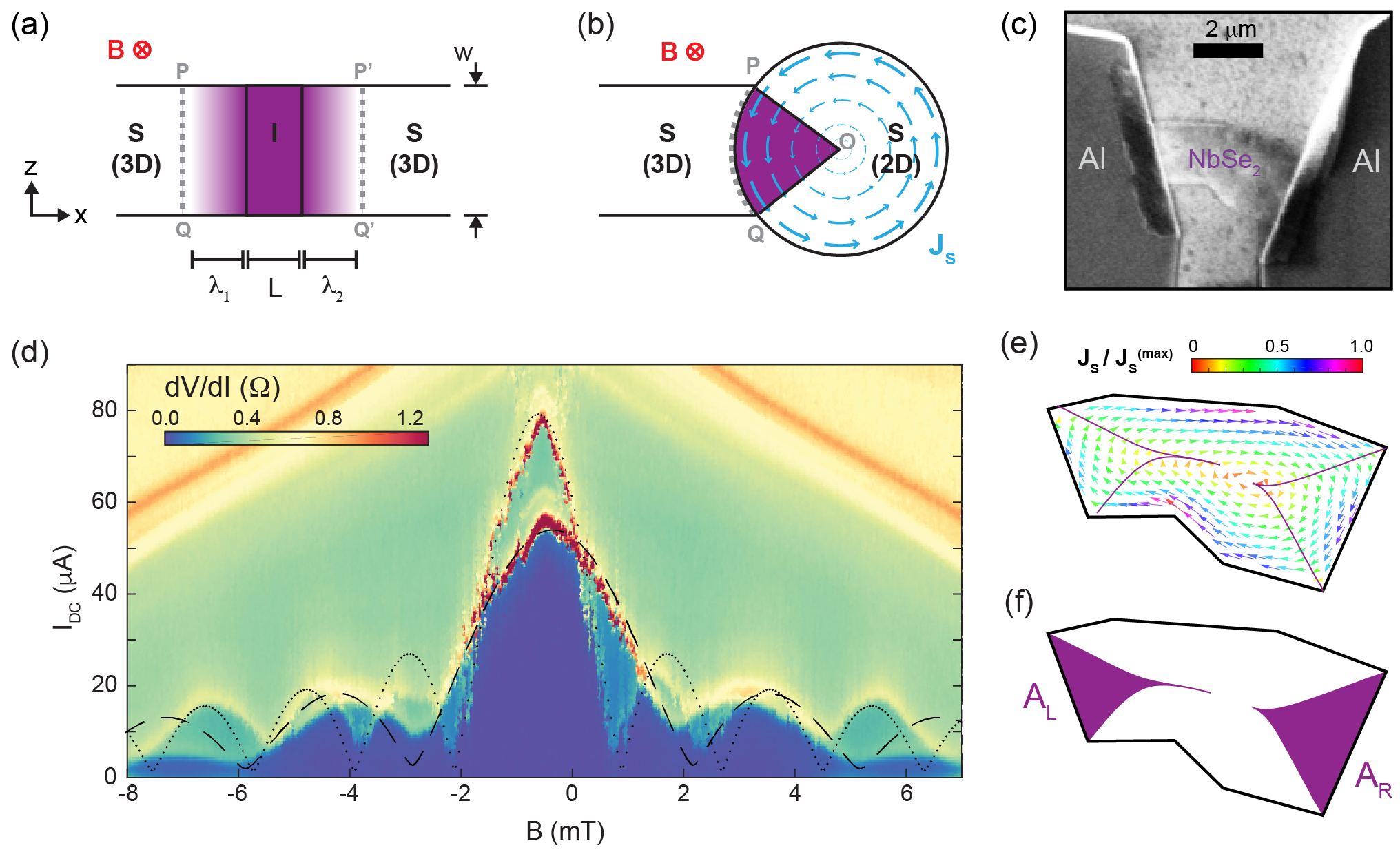} 
\caption{ \textbf{Quantum interference in a two-terminal NbSe$_2$-aluminum device.}
\textbf{(a)}  Model of a 3D-3D Josephson junction. The phase difference $\Delta \phi$ across the junction depends on the $z$-coordinate. $\Delta\phi_{PP'}-\Delta\phi_{QQ'}$ changes by $2\pi$ when one $\Phi_0$ of flux is threaded through the area depicted in magenta. 
\textbf{(b)} Model of a 2D-3D JJ. In response to a field \textbf{B}, circulating supercurrent $\mathbf{J_s}$ (blue arrows) flows in the 2D bulk. The phase difference across the junction winds by $2\pi$ from point $P$ to point $Q$ when one $\Phi_0$ is threaded through the effective area of the 2D-3D junction.  This requires the effective area to be bounded by a contour of constant phase in the interior of the 2D flake, $\overline{POQ}$, perpendicular to $\mathbf{J_s}$ everywhere. \textbf{(c)} SEM image of device F. \textbf{(d)} $dV/dI$ as a function of $I_{\rm DC}$ and $B$ for device F. Two superimposed single-junction critical current vs. field $I_c(B)$ responses are visible. Theoretical $I_c(B)$ curves (black dotted and dashed lines) generated from the numerical simulations shown in E and F are overlaid on the data. \textbf{(e)} Calculated supercurrent distribution $\mathbf{J_S}$ and \textbf{(f)} effective areas $A_L$ and $A_R$ associated with the left and right contacts.}

\label{fig-twoterm-interf}
\end{center}
\end{figure*}

\begin{figure}[ht!]
  \begin{center}    \includegraphics[width=86mm]{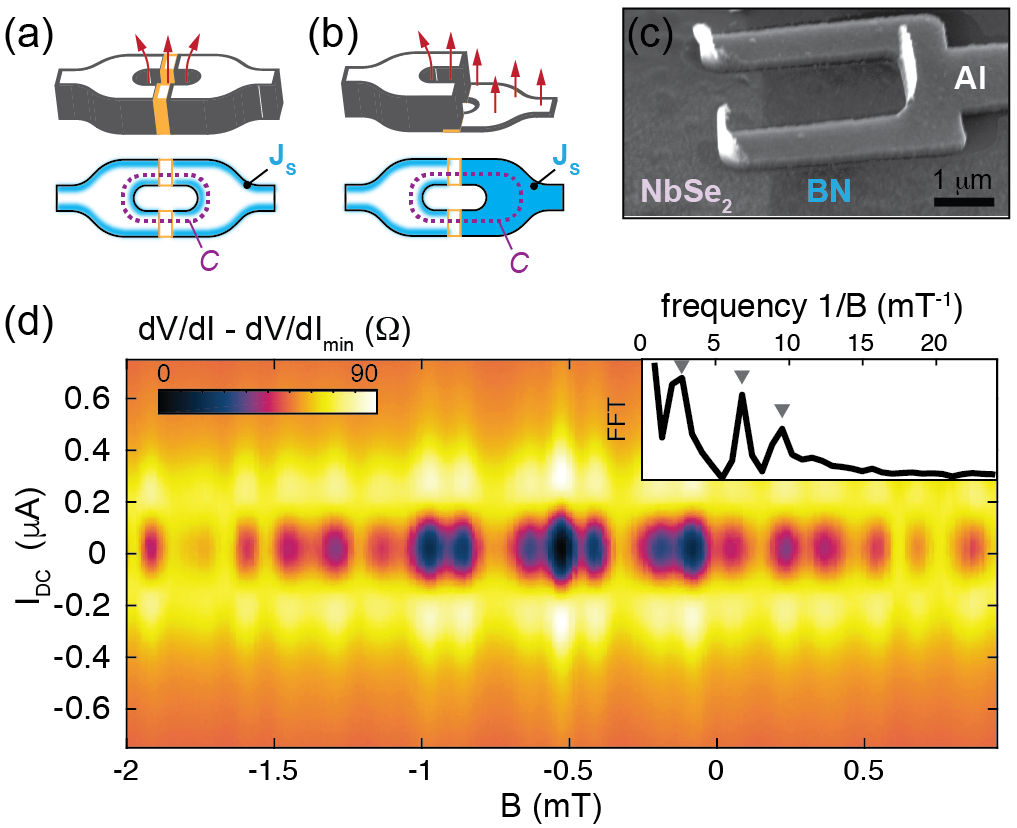} 

\caption{ \textbf{\textbf{Quantum interference in a 2D-3D NbSe$_2$-aluminum SQUID.}}
\textbf{(a)} In a 3D-3D SQUID, the supercurrent $\mathbf{J_S}$ flows in a surface layer of depth $\sim \lambda$. The fluxoid quantization contour $C$ can be chosen deep in the interior of the superconductor such that $\int_C \bf{J_s \cdot d\ell}=0$.  \textbf{(b)} In a 2D-3D SQUID, since $\bf{J_s }$ is non-zero everywhere in the 2D layer, $C$ must be a contour of constant phase inside the bulk, which results in a larger effective area determining the periodicity of $I_c(B)$ oscillations. \textbf{(c)} SEM image of the SQUID ``source'' contact in device C formed by two Al contacts to the \nbse flake. \textbf{(d)} Quantum interference pattern observed in device C. A minimum resistance of $(dV/dI)_{\rm{min}} = \SI{81}{\Omega}$ has been subtracted from the data. Data for SQUID device B (Fig.~\ref{suppfig-othersquiddevice}) shows a true zero resistance as $T\rightarrow 0$. Inset: 
Fast Fourier transform (FFT) amplitude of a linecut along $I_{\rm DC}=0$. The three frequencies, indicated by arrows, are $1/\Delta B_i = $ 2.7, 6.8, and 9.5 mT$^{-1}$ (corresponding respectively to areas $S_i = \Phi_0/\Delta B_i =$ 5.4, 13.6, and 19.1 $\mu \mathrm{m}^2$). The lowest and highest of these areas match the calculated effective areas of the SQUID source and 2D-3D drain contacts.}

\label{fig-squid-main}
  \end{center}
\end{figure}

A schematic of the two-terminal (3D-2D-3D) device is shown in the inset to Fig.~\ref{fig-20}(a). In the pseudo-four-terminal measurement configuration shown, the measured resistance $R(T)$ is the series resistance of the 2D NbSe$_2$ flake, the resistance of the aluminum-NbSe$_2$ interface, and the 3D aluminum leads between the interface and the voltage probes. This measurement configuration allows us to eliminate resistances from the voltage and current leads, but includes the interface resistance between the 3D and 2D superconductors. Fig.~\ref{fig-20}(a) shows a measurement of $R(T)$ between 8 K and 175 mK. The transition at 7 K is that of the NbSe$_2$ flake, whose $T_c$ matches the bulk value of \nbse, reflecting a sample thickness of more than 6 layers \cite{xi_ising_2016}. A second transition occurring at 6.5 K is also associated with the \nbse flake, based on the temperature dependence of the critical current (Fig.~\ref{suppfig-LOGIBR}). A final transition occurs at 1.2 K, which we attribute to the deposited aluminum leads as well as the Al/\nbse contacts. Below this final transition, the total resistance of the device, including the Al-NbSe$_2$ junctions, has dropped to zero within the noise floor of our measurement ($R<10^{-1}\,\Omega)$.  We estimate that $>$60\% of (approximately 8-10) fabricated devices have a resistance of 0.2 $\Omega$ or less in the limit of $T\rightarrow 0$. The remaining devices have residual $T\rightarrow 0$ resistances that range from a few $\Omega$ up to 81 $\Omega$ (see Table~\ref{supptable} for a summary of devices studied).
We have not conclusively identified the origin of the non-zero resistances in some devices, but as we have improved our device fabrication procedures, particularly the in-situ ion milling immediately prior to Al evaporation, the yield of $R=0$ devices has increased and has also resulted in consistently higher critical current densities. We filter each measurement line using both low-pass $RC$ filters ($f_{3\rm{dB}}\approx 10 $ kHz) and radiofrequency filters mounted on the mixing chamber stage of the dilution refrigerator (Fig.~\ref{suppfig-filters}) to reduce the significant effect of current noise \cite{tamir_sensitivity_2019}. 

In Figure \ref{fig-20}B, we plot the differential resistance $(dV/dI)$ as a function of the DC current $I_{\rm DC}$ passing through the device. We observe four critical currents, of which the lower two show periodic modulations with applied magnetic field (Fig.~\ref{fig-twoterm-interf}) while the larger two currents decrease monotonically with applied field (Fig. \ref{suppfig-IBRLOG}). This distinction associates the smaller two critical currents with Josephson junctions' behavior at the Al/\nbse interfaces, and the larger two critical currents with the bulk Al and \nbse. We can further distinguish the bulk Al and \nbse critical currents by their magnetic field and temperature dependences (Fig. \ref{suppfig-IBRLOG} and \ref{suppfig-LOGIBR}).  The two \nbse-Al contacts in this sample are slightly different in size (Table~\ref{supptable}F and Fig.~\ref{fig-twoterm-interf}(c)). We associate the narrower contact with $I_c^{(1)}=\SI{56.5}{\micro A}$ and the wider contact with $I_c^{(2)}=\SI{78.5}{\micro A}$, resulting in critical current densities $J_c^{(1)}=4.4 \times 10^8$\SI{}{A/m^2} and $J_c^{(2)}=4.96 \times 10^8$\SI{}{A/m^2}. 

To gain insight into the nature of the 2D-3D superconducting interface, we studied the response of these two-terminal 3D-2D-3D devices to magnetic flux perpendicular to the plane of the 2D \nbse flake. In Fig.~\ref{fig-twoterm-interf}(d), we plot the differential resistance of such a device as a function of DC current, $I_{\rm DC}$, and magnetic field, $B$. The data appear to display two interference patterns, each with a larger central peak and several smaller satellite peaks. This strongly suggests that we are measuring a superposition of the the individual responses from each junction.  This is supported by our observation that in sequential data sets similar to Fig.~\ref{fig-twoterm-interf}(d), we have observed one set of peaks shifting relative to the other, likely due to the depinning of trapped flux at one of the contacts (Fig.~\ref{suppfig-IBR20260}).

In a 3D-3D superconducting JJ, the critical current $I_c(B)$ is proportional to $|\sin(x)/x|$ with the dimensionless flux $x = \pi BA_{J\!J}/\Phi_0$ penetrating the small area $A_{J\!J} = w\cdot\ell_\text{ef\!f}$ of the barrier region (Fig.~\ref{fig-twoterm-interf}(a)), where $w$ is the width of the junction and $\ell_\text{ef\!f} = L + \lambda_1 + \lambda_2$ is given by the physical length $L$ plus the penetration depths on either side of the junction, $\lambda_1$ and $\lambda_2$ \cite{clarke_squid_2004}. Here, $\Phi_0=h/2e$ is the superconducting flux quantum. In our 2D-3D data (Fig.~\ref{fig-20}D), the two superimposed interference patterns have oscillation periods $\Delta B_i \approx$ \SI{2}{mT} and \SI{4}{mT}, corresponding to JJ areas $A_i = \Phi_0/\Delta B_i$ of 1 \musq and 0.5 \musq respectively.  These numbers are an appreciable fraction of the \nbse flake area and, importantly, are significantly larger than the equivalent interface area produced by our contact method (which, from Fig.~\ref{fig-TEM} and Fig.~\ref{fig-twoterm-interf}(c), has an upper bound of $\sim$0.1 \musq).

In a 2D superconductor of thickness $d \ll \lambda$, where $\lambda$ is the bulk penetration depth, the relevant length scale for the screening of the applied field is the Pearl length $\lambda_{\rm{Pearl}} = 2 \lambda^2/d$ \cite{pearl_current_1964, li_josephson_2019}. $\lambda_{\rm{Pearl}}$ can exceed the size of the sample for very thin flakes and thus the magnetic flux nearly uniformly penetrates the superconductor. In comparison to the 3D-3D JJ, the effective area penetrated by the flux must be much larger in the 2D-3D junction, and the contour around which the superconducting phase winds by 2$\pi$ is determined by the specific flow pattern of the supercurrent $\mathbf{J_s}$ in the 2D flake  (Fig.~\ref{fig-twoterm-interf}(b)). In contrast to 3D London superconductors, the path of this contour in the 2D flake is unique, giving rise to a strong sensitivity of the interference pattern to the precise shape and size of the 2D flake as well as to the placement of the 3D leads.
In order to understand quantitatively the shorter oscillation periods (larger areas) seen in the Fig.~\ref{fig-twoterm-interf} data, we constructed a theoretical model for the flow of supercurrent $\mathbf{J_s}$ in our device, based on numerical solutions of the Ginzburg-Landau equations (see Supplemental Information).  We emphasize that the only input to our numerical model is the geometry of the 2D \nbse flake. The results of the simulation for both contacts are shown in Fig.~\ref{fig-twoterm-interf}(e) and (f), and the $I_c(B)$ for both contacts are superimposed on the data in Fig.~\ref{fig-twoterm-interf}(d). With only the dimensions of the flake as an input parameter, our simple model does a good job of explaining the critical currents: the ratio of the central peak to the satellite peaks is quite accurate, and the predicted areas are within about 8\% of the values measured in the experiment. 
Additionally, our model predicts that the critical currents do not reach zero, a feature of the experimental data as well. Critical current zeroes occur when forward and backward supercurrents across the junction perfectly cancel each other, which can only happen for flakes that have a mirror symmetry with respect to a line perpendicular to the Josephson junction that bisects the junction into two equal segments. As our flakes lack such a symmetry, we do not expect to see critical current zeroes. Agreement between simulation and experiment on this point might be improved if we relaxed our assumption of uniform Josephson coupling along the interface. 

In order to further investigate quantum interference in the presence of near-uniform flux penetration in the 2D bulk, we prepared devices 
consisting of a SQUID-loop ``source'' contact (Fig.~\ref{fig-squid-main}(c)) in series with a 2D-3D ``drain'' contact (Fig. \ref{suppfig-fabanddevices}(h)). We sweep the applied magnetic field and observe a periodic oscillation in the critical current of the SQUID (Fig.~\ref{fig-squid-main}(d)).  A Fourier transform of the data in Fig.~\ref{fig-squid-main}(d) reveals that there are three frequencies $1/\Delta B_i = $ 2.7, 6.8, and 9.5 mT$^{-1}$, which correspond respectively to areas $S_i = \Phi_0/\Delta B_i =$ 5.4, 13.6, and 19.1 \musq. 
None of these precisely matches the physical area of the SQUID shown in detail in Fig.~\ref{fig-squid-main}(c) (7 \musq) and the beating pattern is inconsistent with the ratio of SQUID area to Josephson junction area.

Employing our theoretical model, we calculated the effective areas of both the SQUID source and 2D-3D drain contacts. 
The effective area $S_\text{ef\!f}$ of the SQUID, bounded by the unique contour around which the superconducting phase winds by $2\pi$, is larger than that of the physical area due to the contribution of non-zero $\bf{J_s}$ flowing in the bulk of the 2D flake (Fig.~\ref{fig-squid-main}(b)). We find that the effective SQUID area is $S_\text{ef\!f}=19.7$ \musq, very close to the experimentally determined value of 19.1 \musq calculated from the largest frequency, $1 / \Delta B = 9.5$ mT$^{-1}$ in the Fourier transform.
The drain contact behaves as a 2D-3D Josephson junction with a large effective area, similar to the two-terminal device in Fig.~\ref{fig-twoterm-interf}; our simulation predicts $S_\text{ef\!f}=6.3$ \musq, which corresponds to the lowest observed frequency $1 / \Delta B = 2.7$ mT$^{-1}$ (5.4 $\mu$m$^2$). 
Our observation of a third frequency has no straightforward explanation using our model, but may be related to coupling between phase slips on the two junctions, mixing the frequencies and giving rise to additional sum/difference frequencies. A second SQUID device (C, Fig.~\ref{suppfig-othersquiddevice}) also shows an effective area much larger than the physical loop area of the device.  

In conclusion, we have shown that robust superconducting edge contacts can be made between encapsulated 2D TMD intrinsic superconductors and conventionally evaporated 3D metals. These contacts form Josephson junctions between two dissimilar superconductors, 
whose oscillation period in $I_c(B)$ suggests that the effective areas of the Josephson junctions are significantly larger than the expected areas, an observation which also holds for devices having SQUID geometries. 
The $\sim$\musq effective areas of the Josephson junctions 
suggests that they may be useful in applications involving compact magnetometer arrays or in scanned probes, because of the high ratio of flux-sensitive area to physical device area. These 2D-3D superconducting contacts lay the foundation of a technological pathway to integrate 2D superconductors and heterostructures as components in conventional superconducting circuits. 




Work on device fabrication and measurement by M.R.S. and O.L. was supported by the National Science Foundation PIRE program under award number 1743717. B.M.H. acknowledges support by the Department of Energy under the Early Career award program ({DE-SC0018115}). S.C.B. was supported by the Department of Energy ({DE-SC0018115}) for measurements of the superconducting devices. B.M.H. and S.C.B. acknowledge support in the early stages of this project from the Charles E.~Kaufman Foundation, a supporting organization of The Pittsburgh Foundation, via Young Investigator research grant KA2016-85226. The authors acknowledge use of the Materials Characterization Facility at Carnegie Mellon University supported by grant MCF-677785. This work was performed, in part, at the Nanoscale Fabrication and Characterization Facility, a laboratory of the Gertrude E. and John M. Petersen Institute of NanoScience and Engineering, housed at the University of Pittsburgh. We also thank the Pittsburgh Quantum Institute (PQI) for hosting many productive conversations.


\bibliography{references_Ben}

\onecolumngrid 
\cleardoublepage
\setcounter{figure}{0}
\setcounter{equation}{0}
\renewcommand\thesection{S\arabic{section}}
\renewcommand\thesubsection{S\thesection.\arabic{subsection}}
\renewcommand{\thefigure}{S\arabic{figure}}
\renewcommand{\theequation}{S\arabic{equation}}

\begin{center}
    \textsc{Supplementary Information} for: \\
\textbf{    ``Superconducting Contact and Quantum Interference \\Between Two-Dimensional van der Waals and Three-Dimensional Conventional Superconductors''}
\end{center}

\subsection{Device fabrication}

We fabricate devices by exfoliating $\sim$10-micron-long flakes from a bulk crystal of 2H-\nbse sourced from HQ Graphene. Intrinsically superconducting TMDs are air and water vapor sensitive and begin to degrade in an ambient environment. This necessitates the encapsulation of \nbse by hBN, which we perform using standard dry transfer techniques in an inert N$_2$ glovebox environment to prevent the degredation of the air sensitive \nbse. After a PMMA/MMA bilayer resist mask is patterned using e-beam lithography, an initial etch is made with CHF$_3$/O$_2$ in a reactive ion etching system. The sample is examined and then sealed in an inert environment until it is loaded into a Plassys 8-Pocket e-gun evaporator with ion gun. An initial 3-minute, high-power Ar ion mill step is used to clean and expose fresh surfaces on the exposed cross-sections of the heterostructure, then the substrate is rotated to an angle +/- $30^\circ$ from perpendicular to the direction of evaporation. This angle ensures thorough coverage of the exposed \nbse cross section by the Al, evaporated at 0.3 nm/s. 
Al is evaporated to ensure uniform coverage of heterostructures that range from \SI{30-60}{nm}, devices with leads facing opposite directions have \SI{40}{nm} deposited while the sample is tilted to one side, and then another \SI{40}{nm} of Al deposited while it is tilted the opposite direction.

\begin{figure}[ht!]
  \begin{center}
\includegraphics[angle=0, width=180 mm]{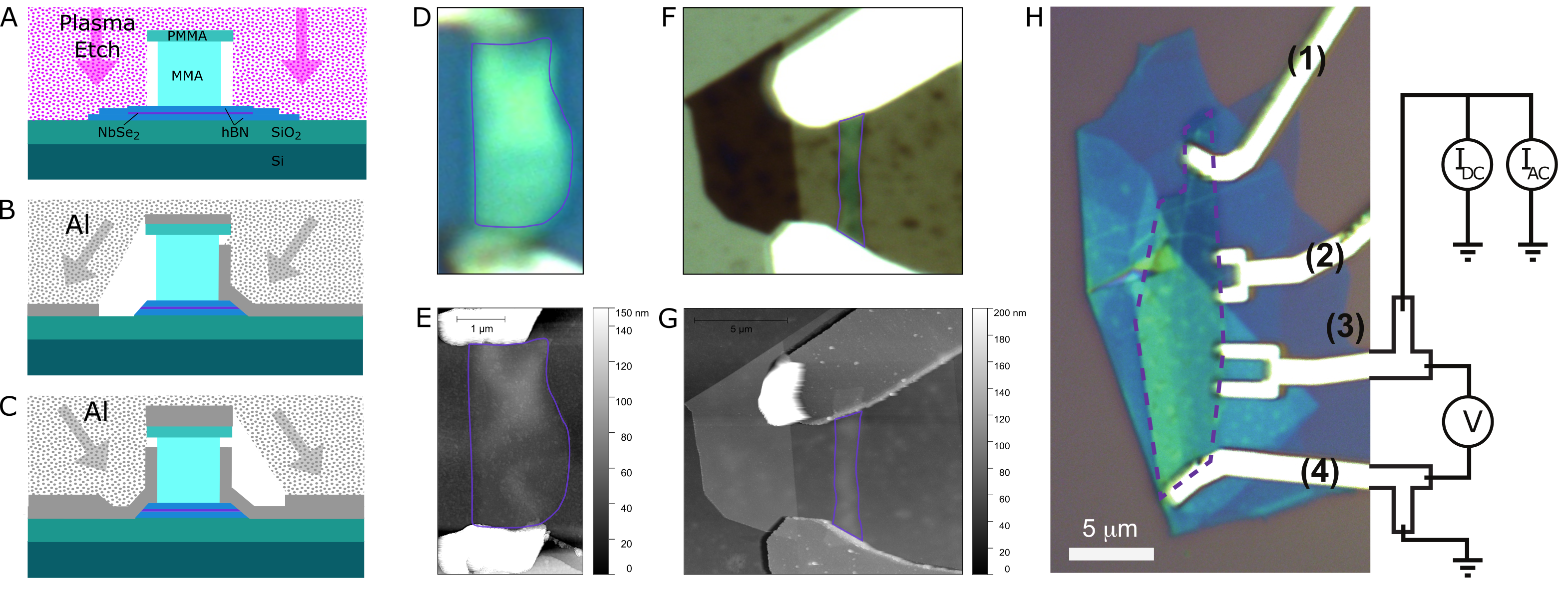} 
\caption{\textbf{(A)} After the lithographic mask is patterned via EBL, a reactive ion etch is used to expose a cross-section of the hBN/\nbse/hBN heterostructure. The stack is then loaded into a Plassys 8 Pocket e-gun Evaporator with ION Gun where a 3 minute Ar ion mill step is used to expose a fresh facet \textit{in situ} before evaporation. \textbf{(B)} Aluminum is evaporated at an angle 30 degrees from vertical. \textbf{(C)} Optionally, additional evaporation angles can be used, but will result in additional aluminum critical currents due to the layering (see Fig.~\ref{suppfig-LinLogA}(b) and (c)). \textbf{(D/E)} Optical and AFM images of sample A. \nbse flake is outlined in purple, scale bar is \SI{1}{\micro m}. \textbf{(F/G)} Optical and AFM images of sample D. \nbse flake is outlined in purple, scale bar is \SI{5}{\micro m}. \textbf{(H)} Optical image and measurement circuit schematic of device C data shown in Fig. \ref{fig-squid-main}}

\label{suppfig-fabanddevices}
  \end{center}
\end{figure}

\begin{table}[ht!]
\centering
\begin{ruledtabular}
\begin{tabular}{cccccccc}
Sample & A & B & C & D & E & F & G \\
 \hline
Geometry & 2 terminal & SQUID & SQUID & 2 terminal & 2 terminal & 2 terminal & 2 terminal \\
Evaporation  & Double angle & Single angle &  Single angle & Single angle &  Single angle & Single angle &  Single angle \\
Thickness & \SI{10}{nm}, \SI{12.5}{nm} & \SI{11.2}{nm} & \SI{11.3}{nm} , \SI{17.5}{nm} & \SI{11.5}{nm} , \SI{12.4}{nm} & \SI{6.2}{nm} & \SI{11.9}{nm} & \SI{8.8}{nm}\\
Contact width & \SI{1}{\micro m}, \SI{1.3}{\micro m} & \SI{1}{\micro m} \slash \SI{1}{\micro m}, \SI{8}{\micro m} & \SI{1}{\micro m} \slash  \SI{1}{\micro m}, \SI{5.2}{\micro m} & \SI{1.3}{\micro m}, \SI{1.4}{\micro m} & \SI{1.5}{\micro m}, \SI{0.5}{\micro m} & \SI{1.4}{\micro m}, \SI{2.0}{\micro m} & \SI{1.2}{\micro m}, \SI{1.4}{\micro m} \\
Contact Length & - & - & - & - & - & \SI{92}{\nano m}, \SI{79}{\nano m} & \SI{95}{\nano m} , \SI{130}{\nano m}\\
$I_c$ & \SI{12}{\micro A}, \SI{26}{\micro A}   & \SI{10}{\micro A} , \SI{40}{\micro A} & \SI{0.15}{\micro A} , \SI{3}{\micro A} & \SI{70}{\micro A} & - & \SI{56.5}{\micro A}, \SI{78.5}{\micro A} & \SI{108}{\micro A}, \SI{128}{\micro A}  \\
J$_c$ & \SI{2\times 10^8}{A/m^2} $^\dagger$ & \SI{1 \times 10^8}{ A/m^2} $^\dagger$ & \SI{5.8\times 10^6}{A/m^2} $^\dagger$ & \SI{5.4\times 10^8}{A/m^2} $^\dagger$ & - &  \SI{4.96\times 10^8}{A/m^2}& \SI{9.5\times 10^{8}}{A/m^2} \\
$R_\text{normal}^\text{contact}$ &  \SI{4}{\ohm} & \SI{2}{\ohm} & \SI{130.6}{\ohm} & \SI{12}{\ohm} & - & \SI{2.06, 1.31}{\ohm} & \SI{200}{\ohm} \\
$R_\text{SC}$ & \SI{\leq 30}{m\ohm} & \SI{200}{m \ohm} & \SI{81.1}{\ohm} & \SI{\leq 30}{m\ohm} & -  & \SI{\leq 130}{m\ohm} & 3-pt \\  
A$^\text{physical}_\text{SQUID}$ & N/A & \SI{4}{\micro m^2} & \SI{7}{\micro m^2} & N/A & N/A & N/A & N/A \\
A$^\text{effective}$ & - & \SI{14}{\micro m^2} & \SI{19}{\micro m^2} & - & - & \SI{0.5}{\micro m^2}, \SI{1}{\micro m^2} & - \\

\end{tabular}
\end{ruledtabular}
\caption{ Properties of 7 samples exhibiting superconducting contacts. Lateral dimensions are obtained from optical and AFM images, Thicknesses are obtained by AFM topography or cross-sectional STEM. Critical current densities are calculated from these dimensions. $R_\text{normal}^\mathrm{{contact}}$ is measured in the critical current measurements where the contacts have gone normal, but while the Al and \nbse flake are still superconducting. $R_\text{SC}$ is measured as $T\rightarrow 0$ with no applied DC current or magnetic field. Effective areas are the outputs from the Ginzburg-Landau model. Contact lengths for samples E and F were measured from STEM images of a sample cross section. $J_\text{C}$ entries marked with a $\dagger$ are calculated using a nominal contact length of 100nm.}
\label{supptable}
\end{table}

\subsection{Theoretical model}
In this section we build a model for the dependence of the Josephson current between the aluminum leads and the NbSe$_2$ flake, similar in spirit to Ref.~\citep{pekker_operation_2005}. We make the assumption that the Josephson energy $E_J$ of the contact between the leads and the flake is much lower than the energy of the screening supercurrents induced by the magnetic field in the flake.  For a JJ with $I_c = \SI{75}{\micro A}$,  $E_J = \Phi_0I_c/2\pi \approx 150$ meV.  The energy of the screening supercurrents is roughly two orders of magnitude higher, with the reasonable assumptions that the (2D) superfluid density is $n_s \sim 10^{15}\,\, \mathrm{cm}^{-2}$ and that the phase gradient in the sample is of the order of $2\pi/L$, where $L\sim \SI{2}{\micro m}$ is a linear dimension of the 2D sample. 
This assumption significantly simplifies our analysis as we can first build a model of the supercurrents in the flake and then use the output of this model to obtain the Josephson currents.

Our starting point for describing the supercurrents in the flake is the Ginzburg-Landau equation
\begin{align}
\alpha \psi + \beta |\psi|^2 \psi + \frac{1}{2m^*} \left( \frac{\hbar}{i} \vec{\nabla}-\frac{2e}{c}\vec{A}\right)^2\psi=0,
\end{align}
where $\psi$ is the complex order parameter, $\vec{A}$ is the gauge field, $m^*$ is the mass of a Cooper pair, and $\alpha$ and $\beta$ the Landau-Ginzburg parameters. For small superconductors (such that the linear size of the 2D sample is much smaller than the Pearl penetration depth) the magnetic field completely penetrates the superconductor, and hence one can ignore the variations in the magnitude of $\psi$. Therefore, we express the order parameter $\psi$ in terms of a (constant) amplitude $\psi_0$ and a position dependent phase $\phi(\vec{r})$: $\psi(\vec{r})=\psi_0 e^{i\phi(\vec{r})}$. Using this form of the order parameter, the Ginzburg-Landau equation becomes
\begin{align}
\vec{\nabla}^2 \phi(\vec{r}) - \frac{2e}{\hbar c} \vec{\nabla} \cdot \vec{A}(\vec{r}) = 0. \label{eq:LG2}
\end{align}
The supercurrents are given by
\begin{align}
\vec{J}(\vec{r})=\frac{2e \hbar}{m^*}\psi_0^2 \left(\vec{\nabla} \phi(\vec{r})-\frac{2e}{\hbar c} \vec{A}(\vec{r}) \right). \label{eq:J}
\end{align}
To obtain the current distribution in the flake we numerically solve equation Eq.~\ref{eq:LG2} subject to the boundary condition that there is no supercurrent across the boundary, i.e. $\vec{J}.\vec{n}|_\partial=0$ where $\vec{n}$ is the unit normal vector on the boundary. We note that in principle the current across the boundary should be non-zero in the regions of contact between the flake and the aluminum leads. We ignore this contribution following our assumption of weak Josephson currents. We also note that  Eq.~\ref{eq:LG2} (supplemented by the boundary conditions) is linear $\phi(\vec{r}) \propto B_z$, and hence it is sufficient to obtain $\phi(\vec{r})$ for a single value of $B_z$ and then scale the resulting solution.

To obtain numerical solutions, we wrote a Mathematica script which allows us to trace the shape of the flake and convert it to into a partial differential equation for $\phi(\vec{r})$ supplemented by Neumann boundary conditions (we use the Landau gauge $\vec{A}(\vec{r})=B_z x \hat{e}_y$). The computed supercurrents in the flake are plotted in Fig.~\ref{fig-squid-main}. 

To compute the Josephson current between the $i$-th lead and the flake we define the phase
\begin{align}
\chi=\varphi+\int dl \left(\vec{\nabla} \phi(\vec{r})-\frac{2e}{\hbar c} \vec{A}(\vec{r})\right),
\end{align}
where $\varphi$ is the superconducting phase associated with the flake and the integral runs around the boundary of the flake. The Josephson current between the $i$-th lead and the flake is given by 
\begin{align}
J_i = J_\text{c} \int_\text{contact} dl \sin\left(\phi_i - \chi \right),
\end{align}
where $ J_\text{c}$ is the critical current density, $\phi_i$ is the superconducting phase of the $i$-th lead, and the line integral runs over the points at which the lead makes contact with the flake.

\begin{figure}[ht!]
  \begin{center}
\includegraphics[width=0.8\textwidth]{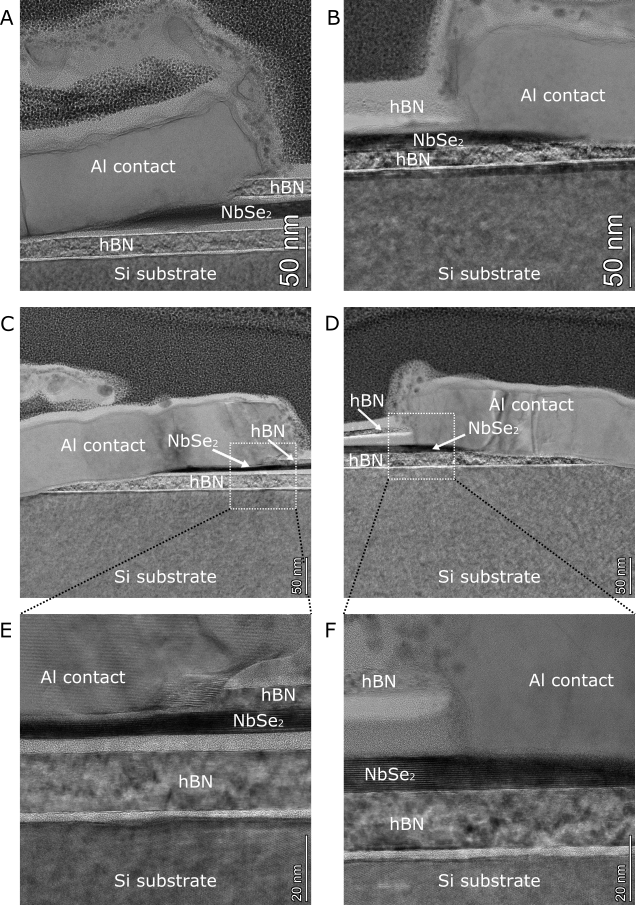} 
\caption{STEM images of the contacts for devices F and G. \textbf{(A)} \SI{56.5}{\micro A} and \textbf{(B)} \SI{78.5}{\micro A} critical current contacts of device F. \textbf{(C)} \SI{128}{\micro A} and \textbf{(D)} \SI{108}{\micro A} critical current contacts of device G. \textbf{(E)}/\textbf{(F)} Detail images of the Al/\nbse contact regions of device G, locations shown in \textbf{(C)}/\textbf{(D)}}

\label{suppfig-TEM}
  \end{center}
\end{figure}

\begin{figure}[ht!]
  \begin{center}
\includegraphics[width=1.0\textwidth]{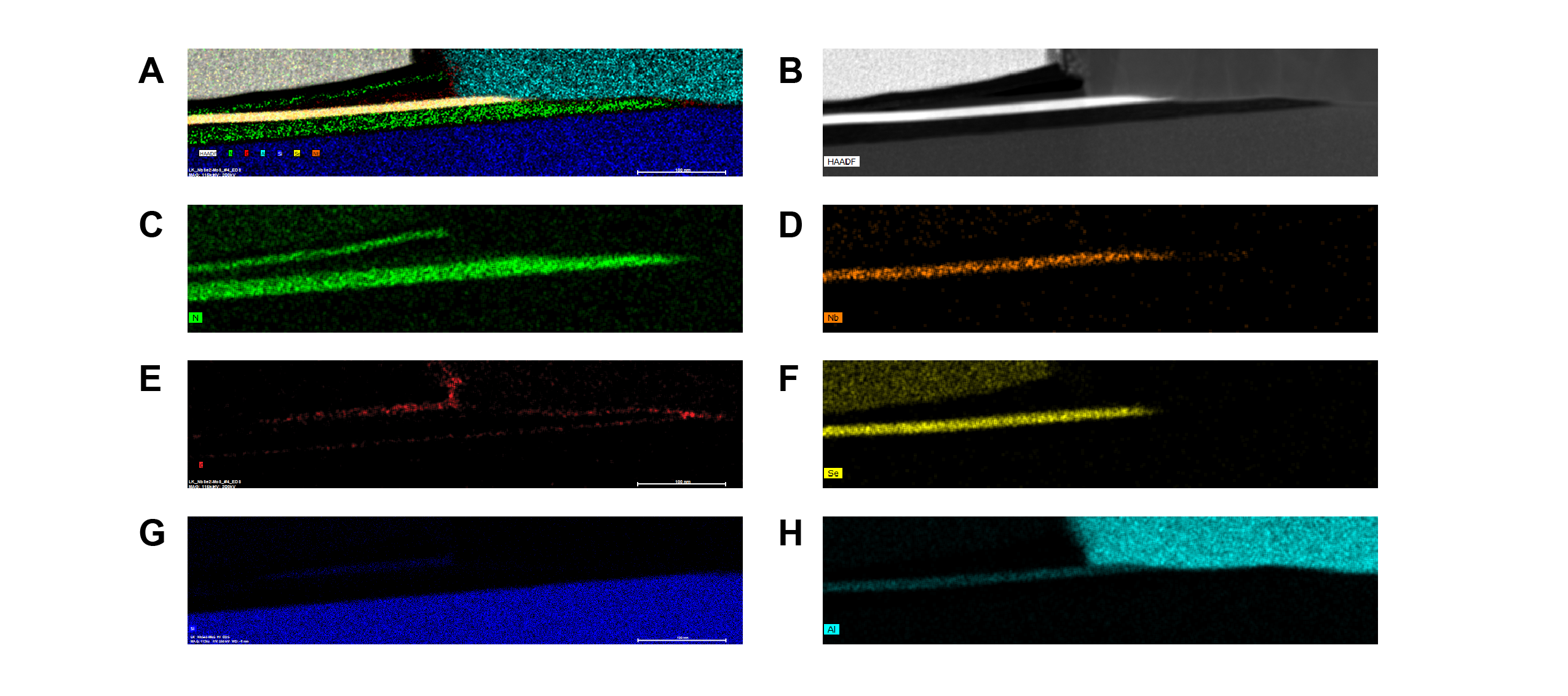} 
\caption{False-color EDS and HAADF maps of device G used to produce Fig.~\ref{fig-TEM}(c).  \textbf{(A)} Composite EDS+HAADF map. Scale bar is 100 nm and is the same for all subfigures. \textbf{(B)} HAADF. \textbf{(C)-(H)} EDS maps of \textbf{(C)} Nitrogen \textbf{(D)} Niobium \textbf{(E)} Oxygen \textbf{(F)} Selenium \textbf{(G)} Silicon, and \textbf{(H)} Aluminum.}

\label{suppfig-EDS}
  \end{center}
\end{figure}

\subsection{STEM specimen preparation and imaging}

Thin lamellae from devices for STEM were prepared using established focused-ion beam lift-out procedures in an FEI (now Thermo Fisher Scientific) Scios focused-ion beam (FIB)/scanning electron microscope (SEM) dual beam system. First, a $\sim$100 nm layer of amorphous carbon of 15 $\mu$m x 2 $\mu$m and then a Pt layer of 15 $\mu$m x 2 $\mu$m x 2 $\mu$m were deposited on the selected contact area using focused ion beam to protect the sample surface from ion damaging during milling; then the surrounding materials around the protected region were milled away using 30 kV high current ion beam; after initial clning and cut, the specimen lamellae were transferred to Cu TEM grid; the lamellae were further thinned to $<$100 nm and further cleaned using 5 kV and 2 kV low-current ion beam. The final lamellae were electron transparent.

Characterization of microstructures and elemental distribution were carried out using a Thermo Fisher Scientific (formerly FEI) Titan Themis 200 G2 probe aberration-corrected field-emission transmission electron microscope equipped with a SuperX EDS (X-ray energy dispersive spectrometer) and operated at an accelerating voltage of 200 kV. STEM images were acquired at a convergence angle of 21.5 mrad with a Fischione HAADF-STEM detector. Drift correction was applied to correct for specimen drift during EDS mapping collection.

\subsection{Magnetotransport measurements}
Magnetotransport measurements were performed using standard low-frequency AC lock-in techniques with an SR860 lock-in amplifier and a Keithley 2400 SMU. The samples were measured in a dilution refrigerator to a minimum temperature of \SI{40}{mK} in fields up to 5T. Fig.~\ref{fig-20}c shows the measurement setup in which each superconducting lead that connects to the \nbse flake splits in two, enabling a 4-pt resistance measurement that would isolate any remaining contact resistance after both the Al lead and \nbse flake have transitioned to superconducting states. A series of filters were used to isolate the sample from high frequency noise, including a low-pass RC filter, Cu tape GHz filters  and a Cu clamshell enclosure, serving as a Faraday cage to encase and isolate the sample\cite{spietz_twisted_2006, chong_transmission_2010,bluhm_dissipative_2008,mandal_efficient_2011,thalmann_comparison_2017}. These filters block higher frequency noise that drives the contact out of the superconducting state, creating a residual resistance \cite{spietz_twisted_2006}. 
dV/dI measurements were taken by the sweeping the sourced DC current at stepped field intervals to form the colorplot in Fig.~\ref{fig-squid-main}(b). 

\begin{figure}[ht!]
  \begin{center}
\includegraphics[angle=0, width=160 mm]{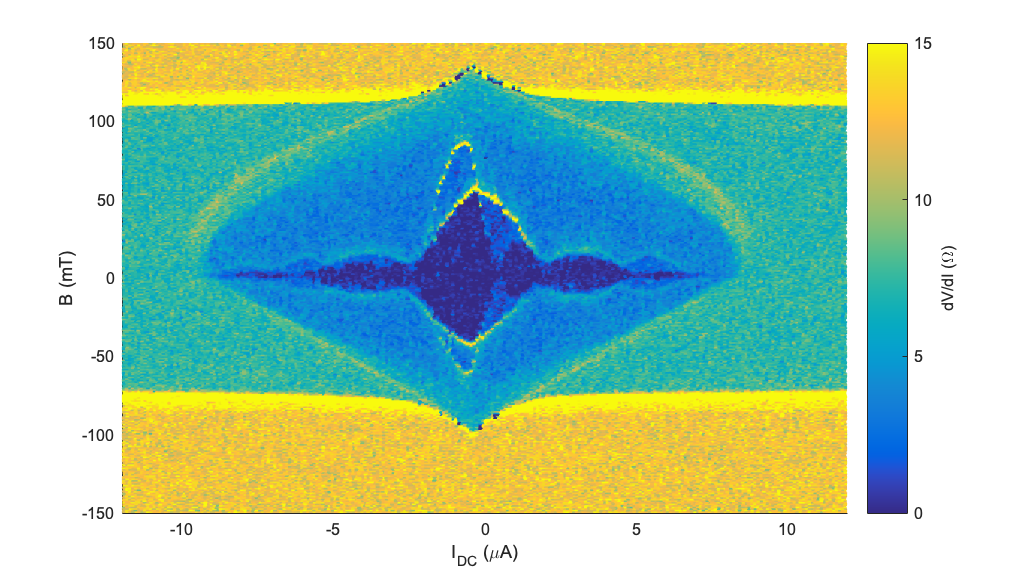} %
\caption{ $dV/dI$ vs.~$I_{\rm DC}$ and $B$ (Device F), with larger $I_{\rm DC}$ and $B$ ranges than in Fig.~\ref{fig-twoterm-interf}(a). $I_{\rm DC}$ was swept from negative to positive, resulting in the top/bottom asymmetry. The critical current of the aluminum leads splits at magnetic fields greater than \SI{2}{mT}, with one component persisting at B fields much greater than the critical field of bulk Al. We attribute this persistent critical current with the narrow regions of the Al leads, where orbital pair breaking effects are slightly suppressed due to the lateral dimensions of the leads ($\sim\SI{1}{\micro m}$). An offset between the centers of the Fraunhofer patterns may indicate that there is a difference in the amount of trapped flux within the regions associated with each contact}

\label{suppfig-IBR20260}
  \end{center}
\end{figure}

\begin{figure}[ht!]
  \begin{center}
\includegraphics[angle=0, width=170 mm]{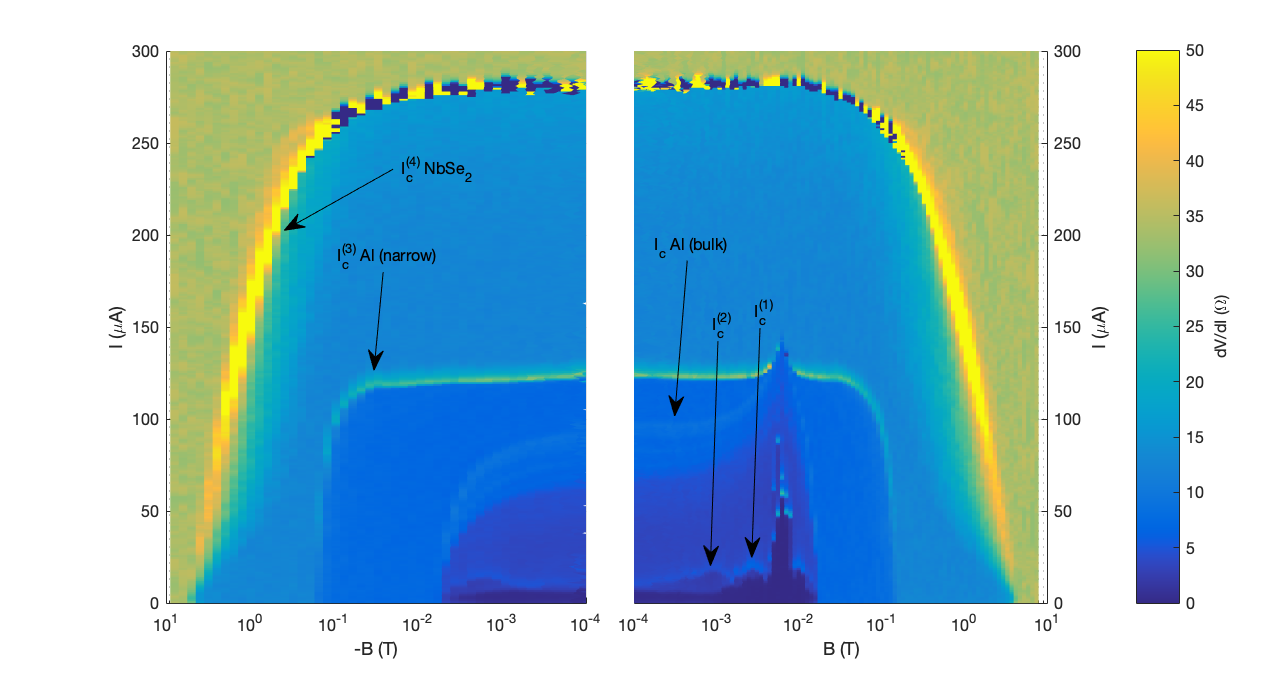} 
\caption{Plot of $dV/dI$ vs.~$\pm$ log($B$) and $I_{\rm DC}$ (Device F). Negative B values are shown on the left in a reversed semilogarithmic plot, leaving a discontinuity between $\pm \SI{0.1}{mT}$. The center of the Fraunhofer patterns is shifted from zero applied field by \SI{5}{mT} due to trapped flux, and all critical fields and currents can be observed: Al and Al/\nbse at $H_{c} \approx \pm \SI{10}{mT}$, narrow Al leads at an elevated $H_{c} \approx \SI{150}{mT}$ due to lateral suppression of orbital pair breaking effects, and the \nbse flake at  $H_{c2} \approx \SI{4.5}{T}$}

\label{suppfig-IBRLOG}
  \end{center}
\end{figure}


\begin{figure}[ht!]
  \begin{center}
\includegraphics[angle=0, width=160 mm]{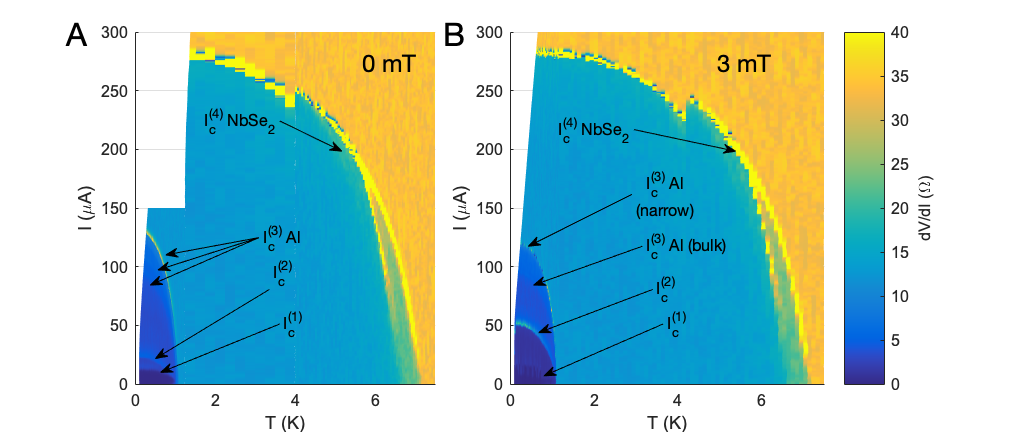} 
\caption{\textbf{(A)} Plot of $dV/dI$ vs.~$I_{\rm DC}$ and $T$ at 0mT (Device F). The critical currents associated with the contacts are well below their maximum at zero flux, indicating that this measurement with zero applied field is measuring a finite amount of trapped flux through the \nbse flake in the areas associated with each junction.
\textbf{(B)} Plot of $I_{\rm DC}$ and $T$ vs.~$dV/dI$ at 3mT. The smaller of the critical currents (associated with the narrower of the 2 contacts) exhibits hopping behavior between two discrete flux states. This suggests that a single vortex is moving in and out of the area of the \nbse flake measured by this contact. The shift in $I_c$ between the two levels is significant. 
In both plots, the $T_c$ of all critical currents associated with the aluminum leads and Al/\nbse contacts is \SI{1.2}{K}. Above \SI{5.5}{K} the \nbse critical current splits into two critical currents with $T_c$ = \SI{6.5}{K},\SI{7}{K}
Note: Discontinuity in $I_c^4$ (\nbse) at 4K is attributed to the different heating behavior of the cryogen free dilution refrigerator above and below 4K}

\label{suppfig-LOGIBR}
  \end{center}
\end{figure}

\begin{figure}[ht!]
  \begin{center}
\includegraphics[width=150 mm]{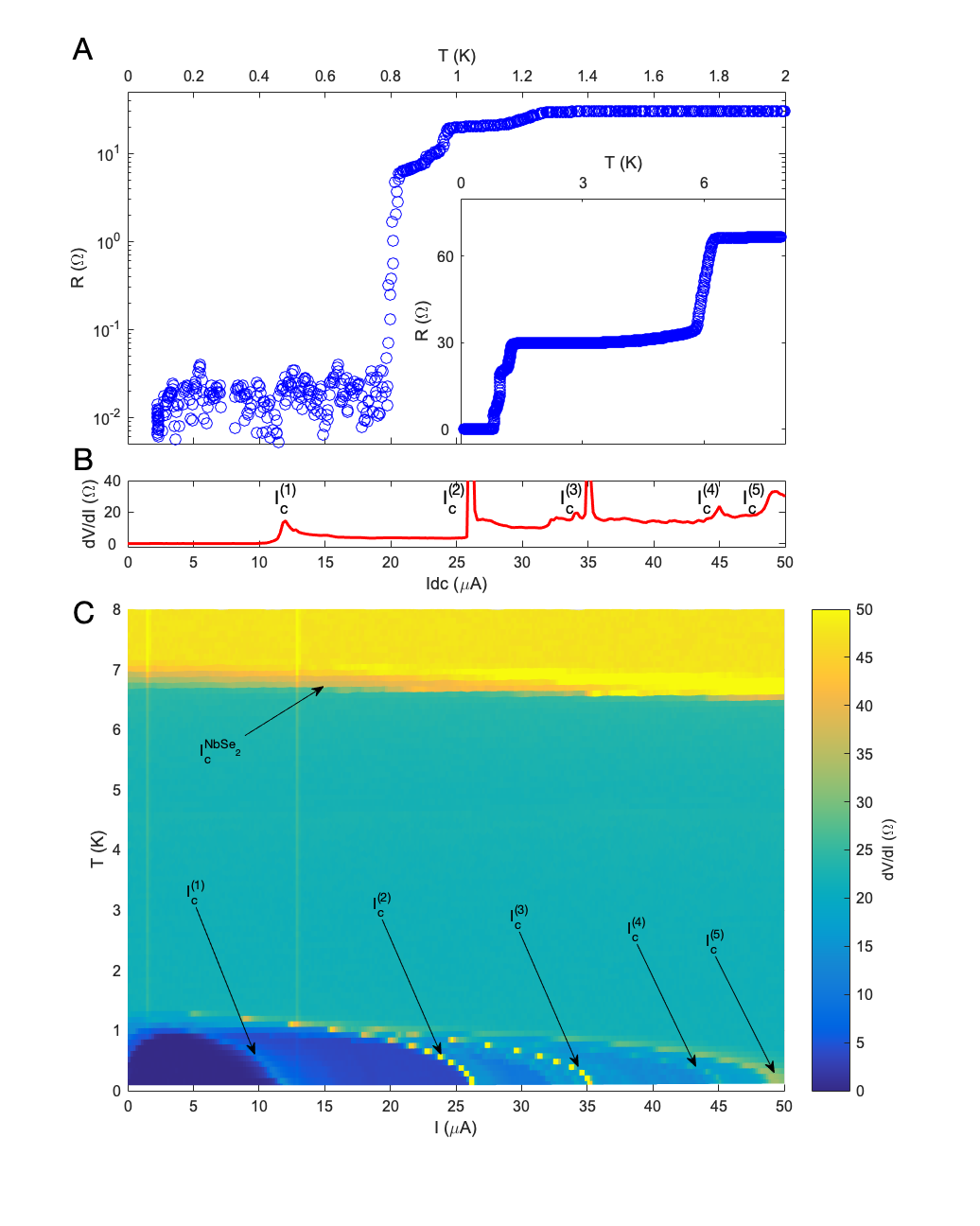} 
\caption{Zero-resistance contact between few-layer NbSe$_2$ and aluminum in Device A.
\textbf{(A)} Semilogarithmic plot of $R(T)$ showing a zero-resistance state limited by the noise floor of roughly 30 m$\Omega$, in sample A. Inset (bottom-right): linear plot of R(T) showing \nbse and multiple Al and Al/\nbse transitions. \nbse is approximately $\SI{1.5}{\micro m}\times \SI{4}{\micro m}$ and \SI{12}{\nm} thick (see Table~\ref{supptable}). \textbf{(B)} Differential resistance $(dV/dI)$ vs. DC current $I_{\rm DC}$ of same device, showing five DC critical currents 1,2,3,4,5 with $I_c \approx 12,26,35,45,49\,\mu $A. The temperature dependence of these critical currents in Fig.~\ref{suppfig-LinLogA}(c) associates them with aluminum and the Al-\nbse contacts instead of the \nbse flake, the critical current of which is not shown. \textbf{(C)} $I_{\rm DC}$ and $T$ vs.~$dV/dI$ shows that all five critical currents observed in B are  associated with the Al/\nbse contacts and Al leads, as evidenced by $I_c^{(1-5)}\rightarrow 0$ at $T_c^{Al}=1.2K$, while $I_c^{\nbse}\rightarrow 0$ at  bulk  $T_c^{\nbse} = \SI{7}{K}$}

\label{suppfig-LinLogA}
  \end{center}
\end{figure}



\begin{figure}[ht!]
  \begin{center}
\includegraphics[width=3.375in]{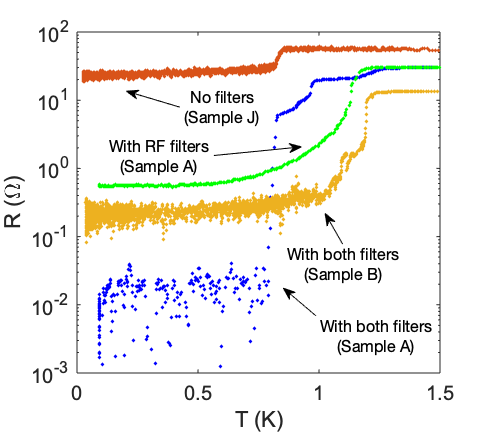}
\caption{Comparison of \textit{R(T)} plots of devices A and B measured with filters, and device J (not listed in Table~\ref{supptable}) without any filters used in the measurement. The curve from Fig.~\ref{suppfig-LinLogA}(a) is reproduced and compared with a measurement on a the same sample using only RF filters with a cutoff frequency in the GHz regime. $R(T)$ for the SQUID sample B with all filters in place shows a residual resistance two orders of magnitude lower than that of Sample J without filters, but unlike Sample A, the resistance remains non-zero.}

\label{suppfig-filters}
  \end{center}
\end{figure}

\begin{figure*}[ht!]
  \centering
\includegraphics[width=175 mm]{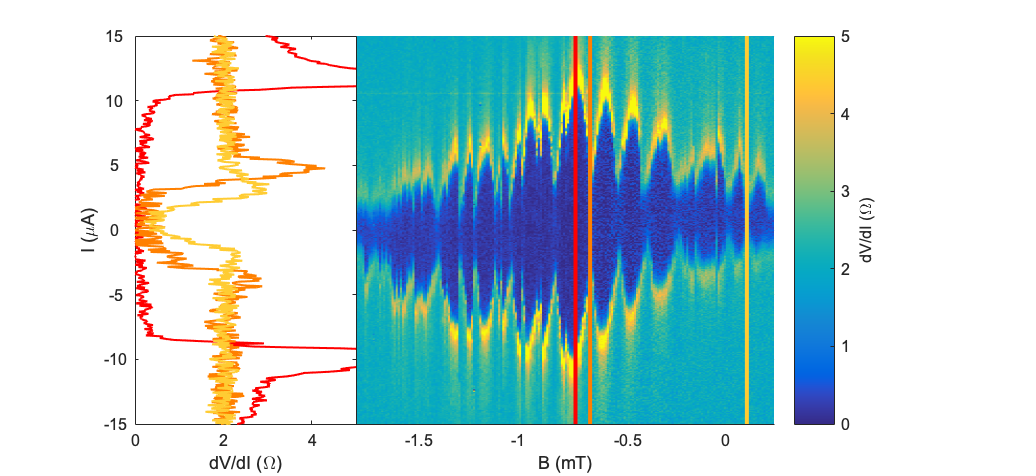} 

\caption{ $dV/dI$ vs.~$I_{\rm DC}$ and $B$ (device B), with several representative linecuts. This SQUID device shows a zero resistance state, unlike the quantum interference pattern shown in \ref{fig-squid-main}B where a residual resistance was subtracted out. The pattern is offset from zero applied field and multiple discontinuities in the interference pattern are present between \SI{-0.7}{mT} and \SI{-1.6}{mT}, providing evidence of vortex motion in and out of the region inside the \nbse flake associated with the flux response of the SQUID. }


\label{suppfig-othersquiddevice}
\end{figure*}

\end{document}